\documentclass[preprint,showpacs,preprintnumbers,amsmath,amssymb]{revtex4}
\usepackage{graphicx}% Include figure files
\usepackage{dcolumn}% Align table columns on decimal point
\usepackage{bm}% bold math
\usepackage{color}

\begin{document}
\title{Formation and migration of native defects in NaAlH$_4$}
\author{Gareth B. Wilson-Short}
\author{Anderson Janotti}
\author{Khang Hoang}
\author{Amra Peles}
\altaffiliation[Present address: ]{United Technologies Research Center,
411 Silver Lane, MS 129-90, East Hartford, CT 06108}
\author{Chris G. Van de Walle}
\email[Corresponding author. E-mail: ]{vandewalle@mrl.ucsb.edu}
\affiliation{Materials Department, University of California, Santa Barbara, CA 93106-5050}

\date{\today}

\begin{abstract}
We present a first-principles study of native defects in NaAlH$_4$. Our analysis indicates that the structure and energetics of these defects can be interpreted in terms of elementary building blocks, which include $V_{\rm{AlH_4}}^+$, $V_{\rm{Na}}^-$, $V_{\rm{H}}^+$, H$_i^-$, and (H$_2$)$_i$. We also calculate migration barriers for several key defects, in order to compare enthalpies of diffusion to experimentally measured activation energies of desorption. From this, we estimate activation energies for diffusion of defects and defect pairs. We suggest that $V_{\rm{AlH_4}}^+$ and H$_i^-$, or $V_{\rm{Na}}^-$ and $V_{\rm{H}}^+$, may be responsible for diffusion necessary for desorption. We discuss the possible role of $V_{\rm{H}}^+$-H$_i^-$ complex formation. The values we find are in the range of activation energies reported for catalyzed desorption.
\end{abstract}


\maketitle

\section{Introduction}

\begin{figure}
\includegraphics[width=3in]{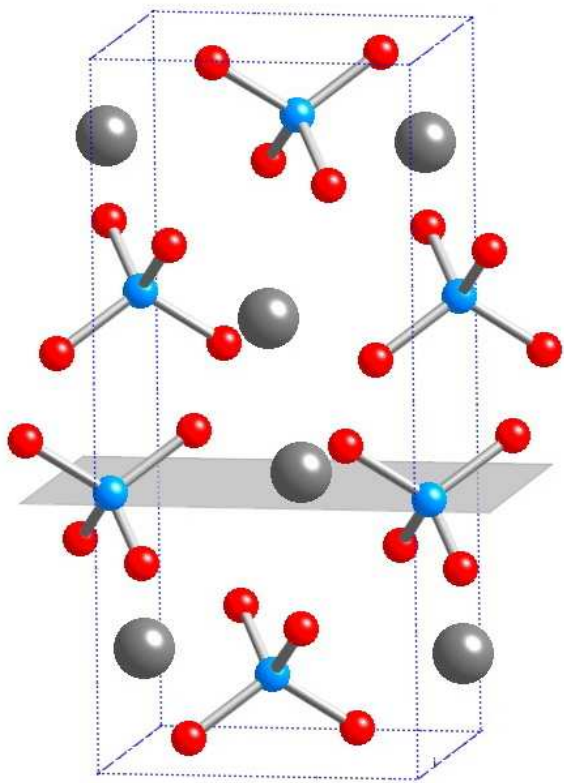}
\caption{(Color online) The body-centered tetragonal structure of NaAlH$_4$. Sodium is represented by large grey spheres, aluminum (at center of tetrahedra) is small dark blue, and hydrogen is small red. A [001] plane is drawn in grey in order to help illustrate depth.\label{bulk}}
\end{figure}

NaAlH$_4$ is an interesting hydrogen storage material. While its theoretical hydrogen capacity by weight (5.6 \%) is not sufficient for automotive applications, it may be useful in other applications. More importantly, as one of the most widely studied hydrogen storage materials, it serves as a prototype for fundamental investigations of kinetics. The structure is shown in Fig.~\ref{bulk}. It can most easily be understood as an NbP ordering of Na$^+$ and (AlH$_4$)$^-$ tetrahedra. Under practical conditions, NaAlH$_4$ undergoes a two-step reaction to release hydrogen:
\begin{equation}
\label{h4-h6}
\rm{NaAlH_4 \rightarrow \frac{1}{3} Na_3AlH_6 + \frac{2}{3} Al + H_2}
\end{equation}
\begin{equation}
\label{h6-h}
\rm{ \frac{1}{3} Na_3AlH_6 \rightarrow NaH + \frac{1}{3} Al + \frac{1}{2} H_2} .
\end{equation}

Reversible absorption and desorption at reasonable temperatures was first accomplished by Bogdanovi{\'c} and Schwickardi in 1997 by adding a few percent titanium.~\cite{97-bogdanovic-tidope} The mechanism of this kinetic improvement has remained controversial. Recent first-princples calculations suggest that titanium may play a role as an electronically active impurity, promoting the diffusion of hydrogen.~\cite{07-peles-theory2} A number of kinetic experiments have been performed on the above reactions,~\cite{02-sandrock-exp, 03-kiyobayashi-exp, 04-gross-exp, 05-kircher-exp, 07-fichtner-exp} and these studies are in reasonable agreement with regards to the reported activation energy of desorption. Desorption of hydrogen and decomposition of NaAlH$_4$ requires not only mass transport of hydrogen but also of aluminum and/or sodium.~\cite{00-gross-exp} This process is likely to be mediated by native defects. Lohstroh and Fichtner suggested that desorption from NaAlH$_4$ is rate-limited by diffusion.~\cite{07-fichtner-exp} Diffusion of native defects was recently approached by Gunaydin {\it et al.} through first-principles methods.~\cite{08-ozolins-pnas} Here we devote further study to this topic.

In this paper we investigate structure and stability of native defects in NaAlH$_4$ based on first-principles density-functional theory. For relevant defects, migration enthalpies are also calculated. These allow us to estimate diffusion activation energies for the various defects that may be responsible for mass transport. We find that most of the relevant defects exist in charge states other than neutral, and that consideration of these charge states is essential for a proper description of migration and kinetics. Section \ref{sec:Methods} describes the computational approach. In Sec.~\ref{sec:results} we report our results. Section \ref{sec:Discussion} contains a discussion and comparison with experiment.

\section{Methods}
\label{sec:Methods}

We use the Vienna Ab-Initio Simulation Package (VASP)~\cite{vasp1, vasp2, vasp3} to perform density functional theory (DFT) calculations within the Perdew-Berke-Erzerhoff (PBE)~\cite{pbe96} generalized gradient approximation (GGA). The plane-wave cutoff is 500 eV, and projector-augmented wave potentials~\cite{paw} are used. In the case of sodium, 2{\it p} orbitals are included in the valence description of the atom. Our calculated theoretical lattice parameters for NaAlH$4$ are $a$ = 5.01 \AA~and $c$ = 11.12 {\AA}, within 0.7\% of the experimental values.~\cite{03-hauback-NaAlH4}

We calculate point defects in a supercell containing 96 atoms. The supercell dimensions are kept at the theoretical bulk lattice parameters, but of course the atoms within the cell are fully relaxed. This method has been used previously in the study of sodium aluminum hydride.~\cite{07-peles-theory, 07-peles-theory2} The formation energy of a defect is a key quantity, determining its concentration in the lattice through the relation:~\cite{vdw-rev-04}
\begin{equation}
c(\mathrm{X})= N_{\rm sites}N_{\rm config} \exp[E^f({\rm X})/kT],
\label{eq:conc}
\end{equation}
where $E^f({\rm X})$ is the formation energy of defect ${\rm X}$, $N_{\rm sites}$ is the number of lattice sites per unit volume on which the defect can be incorporated, and $N_{\rm config}$ is the number of configurations per site in which the defect can be formed.

NaAlH$_4$ is an insulator with a wide band gap -- it is expected that native defects exist in charge states other than neutral. There exists a clear prescription for calculating the formation energies of charged defects in a manner that accurately describes the thermodynamic reservoirs for atoms and for electrons.~\cite{vdw-rev-04} The formation energy of a charged defect ${\rm X}$ is calculated as:
\begin{equation}
\label{eq:eform}
E^f({\rm X}^q)=E_{\rm tot}({\rm X}^q)-E_{\rm tot}({\rm bulk})+q \epsilon_F +\sum_{i=0}{n_i\mu_i}.
\end{equation}
E$_{\rm tot}({\rm X}^q)$ is the total energy of a supercell containing the defect in charge state $q$. Similarly, E$_{\rm tot}({\rm bulk})$ is the total energy of a supercell without a defect. The last term in the expression ensures stochiometric balance, with $n_i$ representing the number of atoms exchanged with the reservoir with chemical potential $\mu_i$. Our chemical potentials are referenced to the standard state (i.e., bulk bcc Na, bulk fcc Al, and H$_2$ molecules at $T$=0). Presented separately from the atomic reservoir chemical potentials is $\epsilon_F$, the chemical potential for electrons or Fermi energy. We keep with convention and reference the Fermi energy to the valence-band maximum of the bulk material. In this study, a correction to the formation energy of a charged defect is made through averages of the electrostatic potential in regions far away from the defect.~\cite{vdw-rev-04}

Migration energies were calculated by using an implementation of the Nudged-Elastic Band (NEB) method~\cite{00-jonsson-theory} within VASP. For charged defects, in principle the aforementioned correction related to the electrostatic potential alignment could vary along the diffusion path; we have checked that for singly charged defects this correction term is small. In the case of diffusion of $V_{\rm{Na}}^-$, inclusion of this energetic correction caused a decrease of 0.06 eV in the energy of the saddle point relative to the minima.

The enthalpy of diffusion is the sum of formation and migration enthalpies. These enthalpies do not include any pressure term from gas-phase H$_2$ -- to first approximation this term should drop out of an activation energy measurement. In the following, we therefore focus on activation energies for diffusion that we take to be the sum of calculated formation and migration energies.

\section{Results}
\label{sec:results}

\subsection{Chemical potentials}
\label{sec:chempot}

Our calculated formation energies are completely general, and can be applied to any condition described by a set of atomic chemical potentials. I.e., the atomic chemical potentials $\mu_i$ in Eq.(~\ref{eq:eform}) are variables that can describe different sets of experimental conditions. It is useful to consider various possible scenarios that lead to specific constraints.

Equilibrium with NaAlH$_4$ implies that
\begin{equation}
\label{eq:NaAlH4}
\mu_{\rm Na} + \mu_{\rm Al} + 4 \mu_{\rm H} = \Delta H_f({\rm NaAlH_4}),
\end{equation}
where $\Delta H_f({\rm NaAlH_4})$ is the enthalpy of formation of NaAlH$_4$; our calculated value for this quantity is $-$0.824 eV (experiment: $-$1.205 eV, Ref.~\onlinecite{06-lee-review}).
Similarly, equilibrium with Na$_3$AlH$_6$ implies that
\begin{equation}
\label{eq:Na3AlH6}
3 \mu_{\rm Na} + \mu_{\rm Al} + 6 \mu_{\rm H} = \Delta H_f({\rm Na_3AlH_6}),
\end{equation}
where $\Delta H_f({\rm Na_3AlH_6})$ is the enthalpy of formation of Na$_3$AlH$_6$; our calculated value is $-$1.784 eV (experiment: $-$2.475 eV, Ref.~\onlinecite{06-lee-review}).

The difference between our calculated enthalpies of formation and experiment can be partly attributed to zero-point corrections to the vibrational energies.~\cite{05-tanaka-theory} We do not consider such corrections here; their inclusion would not change our qualitative conclusions for formation energies, since typically significant cancellation occurs between terms in the defect and in the reservoirs.~\cite{vdw-rev-04} I.e., energy {\it differences} between comparable solids are typically well described. We note, for instance, that our calculated heat of reaction for Eq.~(\ref{h4-h6}) is 0.23 eV, and for Eq.~(\ref{h6-h}) it is 0.20 eV; these values are quite close to the experimental values of 0.38 eV and 0.24 eV, respectively.~\cite{06-lee-review}

For calculating defect concentrations, in principle one would have to include the contributions from finite temperature vibrational entropies. A rigorous treatment of this effect requires an evaluation of the vibrational spectrum for each of the defects. This would involve an extraordinary computational effort. Nonetheless, it would not lead to a deeper understanding of the effects discussed here, so we consider this task beyond the scope of the present work. Reasonable estimates indicate that such contributions are relatively small, mainly due to cancellation when comparing the vibration entropy of the defect system with the ideal host and reservoir.\cite{vdw-rev-04} In addition, at the temperatures of interest in the present work (noting that NaAlH$_4$ is stable up to 200$^\circ$C), the $TS$ term will not make a significant contribution to the free energies of formation.~\cite{07-peles-theory}

For purposes of presentation of formation energy results, it is convenient to choose a specific set of chemical potentials that are intended to be close to those relevant for dehydrogenation. We will choose the chemical potentials of Al, Na, and H by assuming equilibrium with Na$_3$AlH$_6$, NaAlH$_4$, and Al. Using our calculated energies, this results in a value for $\mu_{\rm H}$=$-$0.12 eV, quite close to the Gibbs free energy of H$_2$ gas at 1 atm and 303 K, the equilibrium temperature of Na$_3$AlH$_6$, Al, NaAlH$_4$, and H$_2$ at 1 atm.~\cite{06-lee-review} This agreement supports our argument that this set of chemical potentials is representative of real conditions.

While this particular set of chemical potentials presents a convenient and relevant set for presenting our formation energy results, it does not preclude us from examining situations that correspond to different choices of chemical potentials. The corresponding formation energies can easily be derived based on the values given in this paper and the general expression for formation energy. Examples will be given in Sec.~\ref{sec:Discussion}.

\subsection{Hydrogen-related defects}

\begin{figure}
\includegraphics[width=5in]{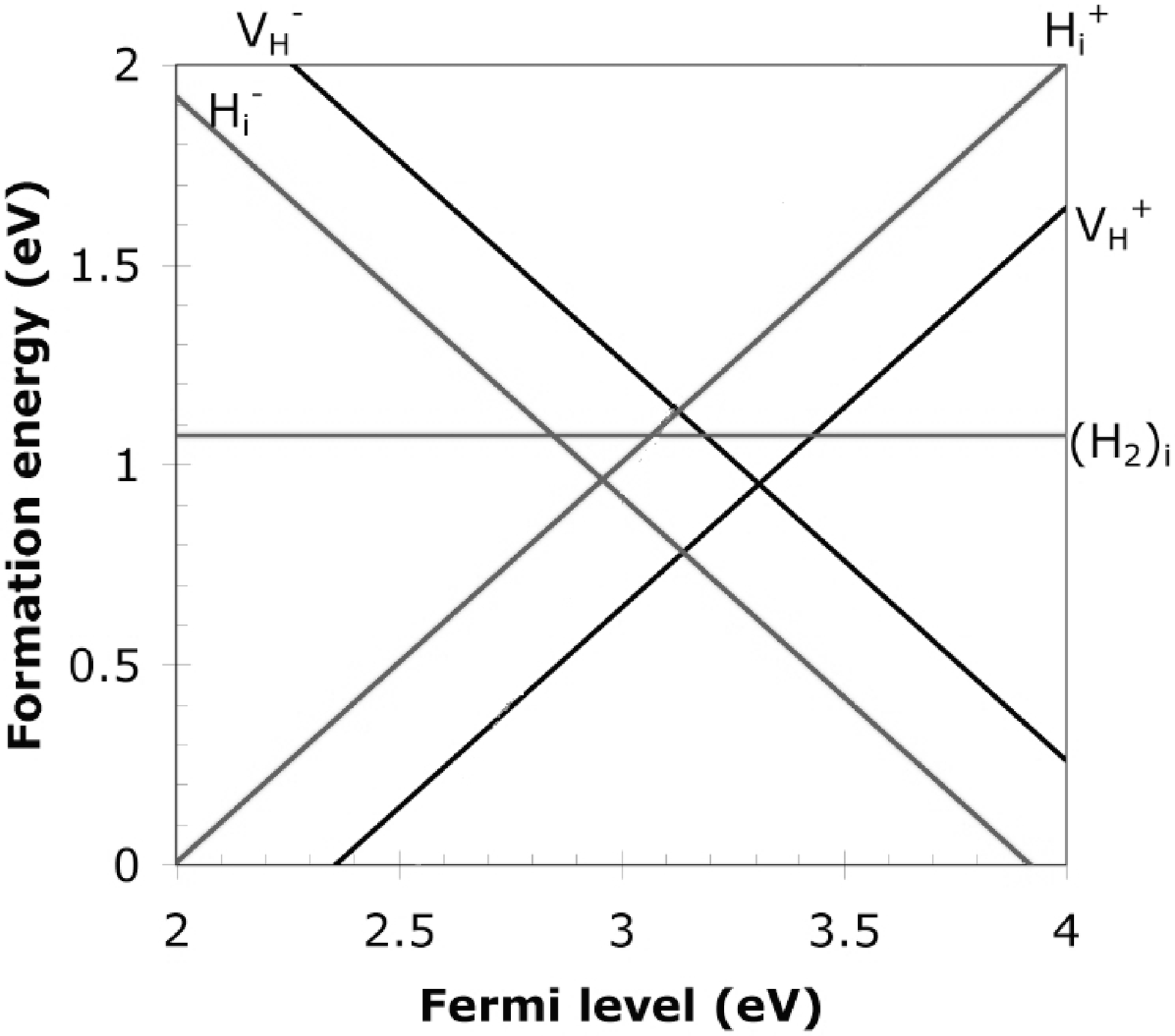}
\caption{Formation energy of hydrogen-related native defects in NaAlH$_4$, plotted as a function of Fermi level with respect to the valence-band maximum. Atomic chemical potentials were chosen to reflect equilibrium with Na$_3$AlH$_6$, NaAlH$_4$, and Al.\label{Hdef}}
\end{figure}

We start by presenting our first-principles results for hydrogen-related defects in NaAlH$_4$ in Fig.~\ref{Hdef}. These values have been calculated before,~\cite{07-peles-theory, 07-peles-theory2} but we include them here for completeness and ease of comparison. Our present numbers are in good agreement with the previous results;~\cite{07-peles-theory, 07-peles-theory2} small differences arise from the use of a different exchange-correlation functional. In agreement with the previous calculations~\cite{07-peles-theory2} we find that the neutral defects, $V_{\rm{H}}^0$ and H$_i^0$, have higher formation energies than the charged defects for all Fermi-level positions. They are therefore not included in Fig.~\ref{Hdef}.

\begin{figure}
\includegraphics[width=5in]{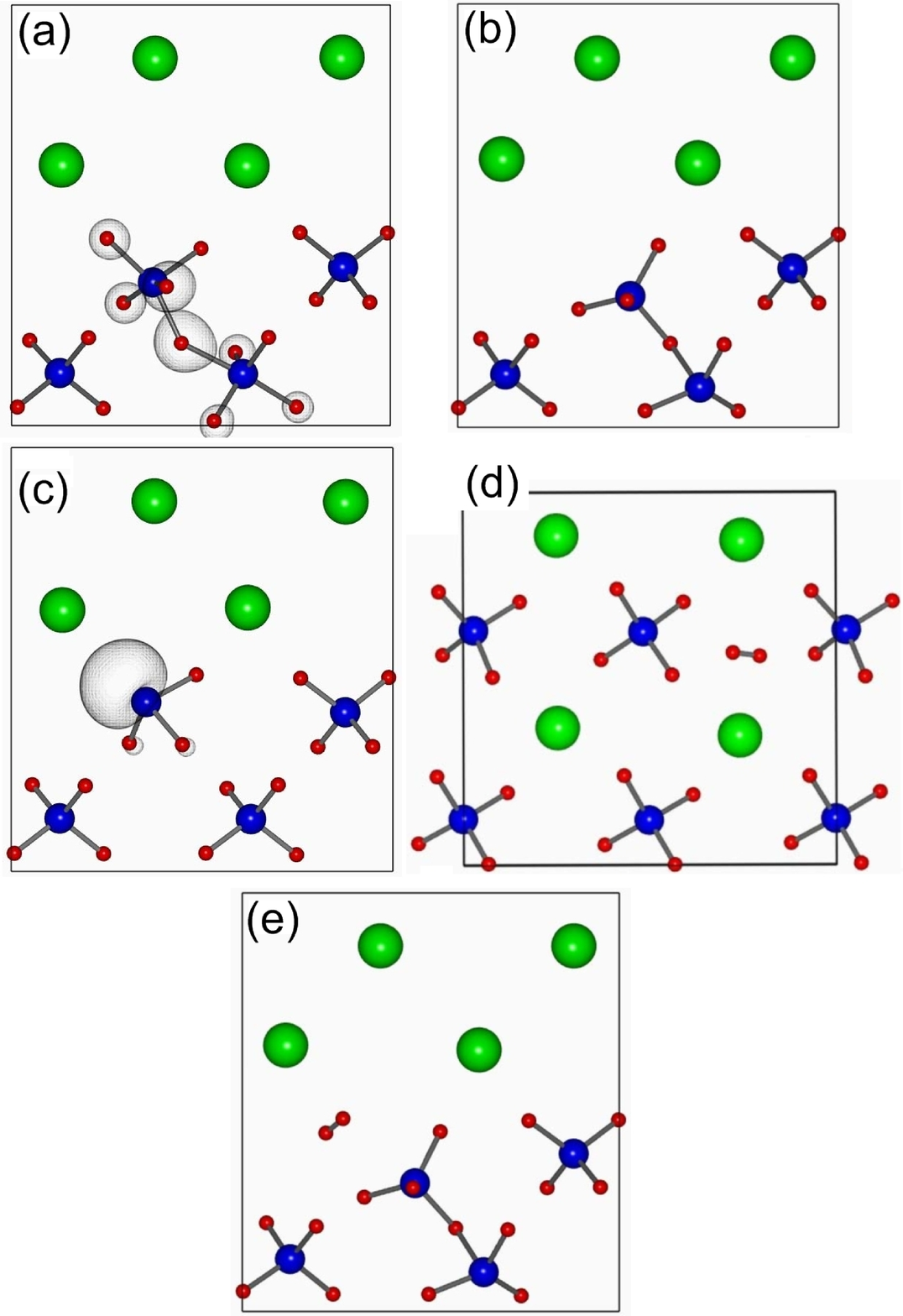}
\caption{(Color online) The structures of (a) H$_i^-$, (b) $V_{\rm{H}}^+$, (c) $V_{\rm{H}}^-$, (d) (H$_2$)$_i$, and (e) H$_i^+$. In the case of H$_i^-$ and $V_{\rm{H}}^-$ an isosurface of the highest occupied orbital is shown, in order to illustrate the defect-character of this orbital. The electron density at the isosurfaces is 0.06 electrons/{\AA}$^3$. All of the structural diagrams are of sodium and aluminum atoms in a [010] plane (and coordinated hydrogens), except for the case of (H$_2$)$_i$ in panel (e) where a [001] plane is shown.\label{def-h}}
\end{figure}

H$_i^-$ can be thought of as the addition of H$^-$ to the system. The structure of H$_i^-$ is shown in Fig.~\ref{def-h}(a). This defect can be viewed as an (Al$_2$H$_9$)$^{3-}$ unit; i.e., it is composed of two (AlH$_4$)$^{-}$ units with an additional H$^-$ located midway. The central hydrogen of this defect structure sits in a ``bridge bond''-type arrangement between two aluminum sites at distances of 1.78 and 1.84 \AA. An isosurface of the charge density associated with the defect state, which occurs at 2.5 eV above the valence-band maximum, is included in Fig.~\ref{def-h}(a). One might expect that, due to Coulomb interaction, H$_i^-$ would prefer to sit next to a Na atom in NaAlH$_4$. Indeed, we found a configuration in which H$_i^-$ is located near two Na, with Na-H distances of 1.95 \AA. However, this configuration is metastable, being 0.6 eV higher in energy than the lowest energy configuration shown in Fig.~\ref{def-h}(a).

The migration path for this defect can be thought of as diffusion of the defect complex along the 4$_1$ screw axis of the system. One of the (AlH$_4$)$^{-}$ units to which the H$_i^-$ is attached slightly rotates and moves closer to an adjacent (AlH$_4$)$^{-}$ until a new bridge bond forms, while simultaneously the bridge bond with the original (AlH$_4$)$^{-}$ is broken. The resulting migration barrier of this defect is 0.16 eV. Such low values indicate a very high diffusivity of the point defect. Note that the migration barrier reported in the present work is slightly lower, by 0.06 eV, than the value reported in Ref.~\onlinecite{07-peles-theory2}. This small difference is attributed a slight difference in the migration paths.

$V_{\rm{H}}^+$ can be viewed as the removal of H$^-$ from the system. In response to this, the resulting undercoordinated Al site shifts towards an adjacent (AlH$_4$)$^-$ tetrahedron. The defect can be thought of as a (Al$_2$H$_7$)$^{-}$ complex [see Fig.~\ref{def-h}(b)]. The migration barrier for this species has previously been calculated to be 0.26 eV.~\cite{07-peles-theory2}

$V_{\rm{H}}^-$ can be thought of as the extraction of a proton from the system. Compared to the large atomic rearrangements observed for the other defects, the formation of $V_{\rm{H}}^-$ has a relatively small impact on the geometry [see Fig.~\ref{def-h}(c)]. A plot of the highest occupied Kohn-Sham state (1.2 eV above the valence-band maximum) shows that an aluminum lone pair has replaced the missing H$^+$. We have calculated the migration barrier for this defect by moving a H atom from a neighboring (AlH$_4$)$^-$ unit into the vacancy. The resulting migration barrier of this defect is 0.92 eV. This barrier is significantly higher than that of the other hydrogen-related defects. The reason is that the saddle-point configuration consists of a hydrogen atom located midway between two AlH$_3$ units. Such a configuration is favorable in the case of a positive charge state (see the case of $V_{\rm H}^+$), but quite high in energy in the case of a negative charge state, due to the fact that an antibonding state resulting from the interaction between the Al atoms needs to be occupied in order to accommodate the charge.

(H$_2$)$_i$ is an interstitial molecule inside the hydride [see Fig.~\ref{def-h}(d)]. Figure~\ref{Hdef} shows that this defect is relatively high in energy. The calculated H-H bond length is 0.76 {\AA}, very close to that calculated for an isolated molecule (0.75 {\AA}). Migration of this defect has a calculated barrier of 0.25 eV. The bond length of the H$_2$ dimer itself is very well preserved along the migration path.

H$_i^+$, finally, has a surprising structure [see Fig.~\ref{def-h}(e)] that can be understood as a complex between $V_{\rm{H}}^+$ and (H$_2$)$_i$, both discussed earlier [see Figs.~\ref{def-h}(b) and (d)]. Within this model, it is interesting to consider the binding energy of this pair. The formation energy of H$_i^+$ is lower than the sum of the formation energies of $V_{\rm{H}}^+$ and (H$_2$)$_i$ by 0.71 eV. This binding energy is independent of any chemical potential for either atomic species or electrons. This relatively large binding energy can be attributed to the unfavorable sterics of the component (H$_2$)$_i$. When (H$_2$)$_i$ is in the presence of a vacancy, the induced local strain in the lattice largely disappears.

\subsection{Aluminum-related defects}

\begin{figure}
\includegraphics[width=5in]{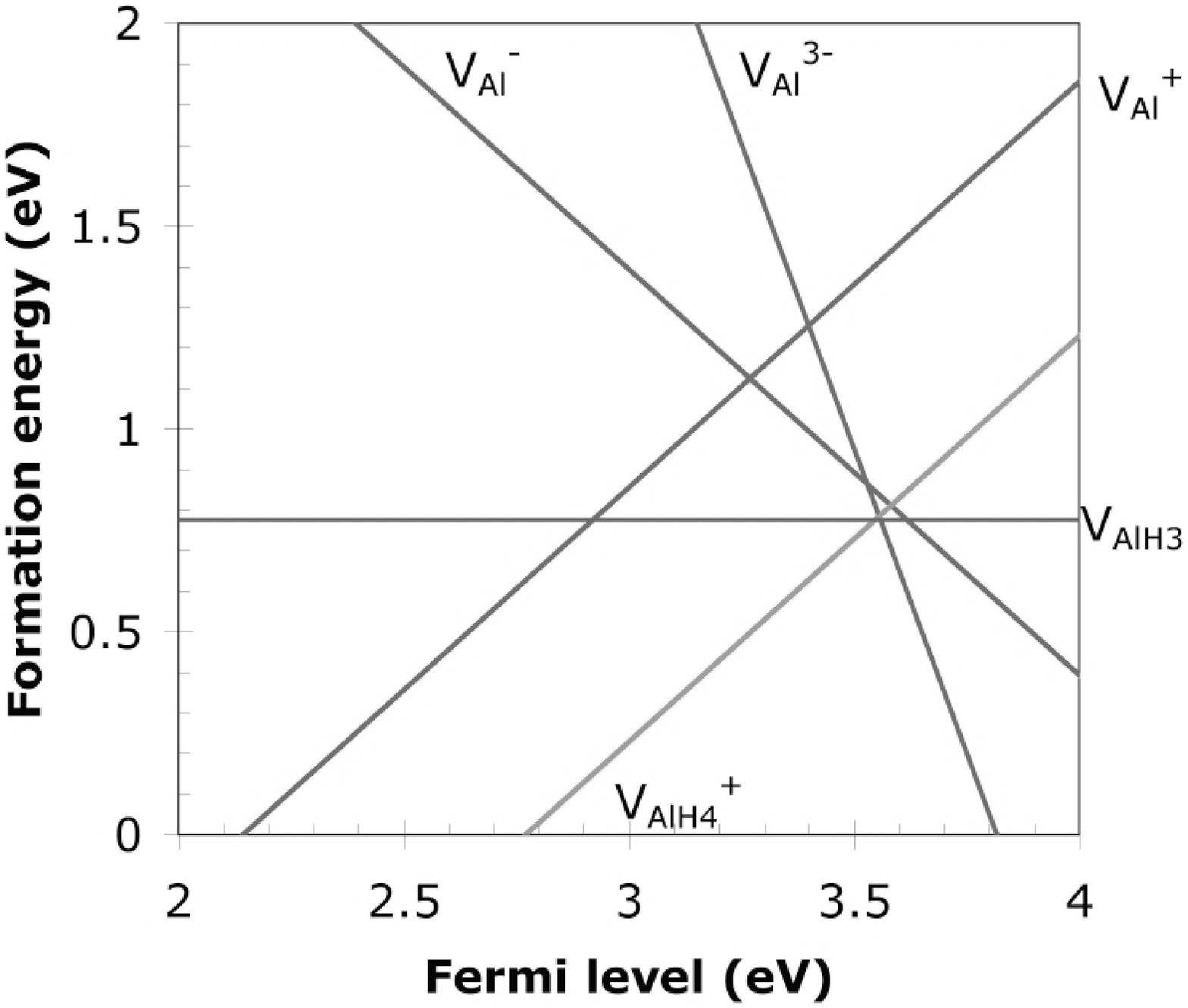}
\caption{Formation energy of aluminum-related native defects in NaAlH$_4$, plotted as a function of Fermi level with respect to the valence-band maximum. Atomic chemical potentials were chosen to reflect equilibrium with Na$_3$AlH$_6$, NaAlH$_4$, and Al.\label{Aldef}}
\end{figure}

Next we turn to native defects related to the Al site. The calculated formation energies are shown in Fig.~\ref{Aldef}.

\begin{figure}
\includegraphics[width=5in]{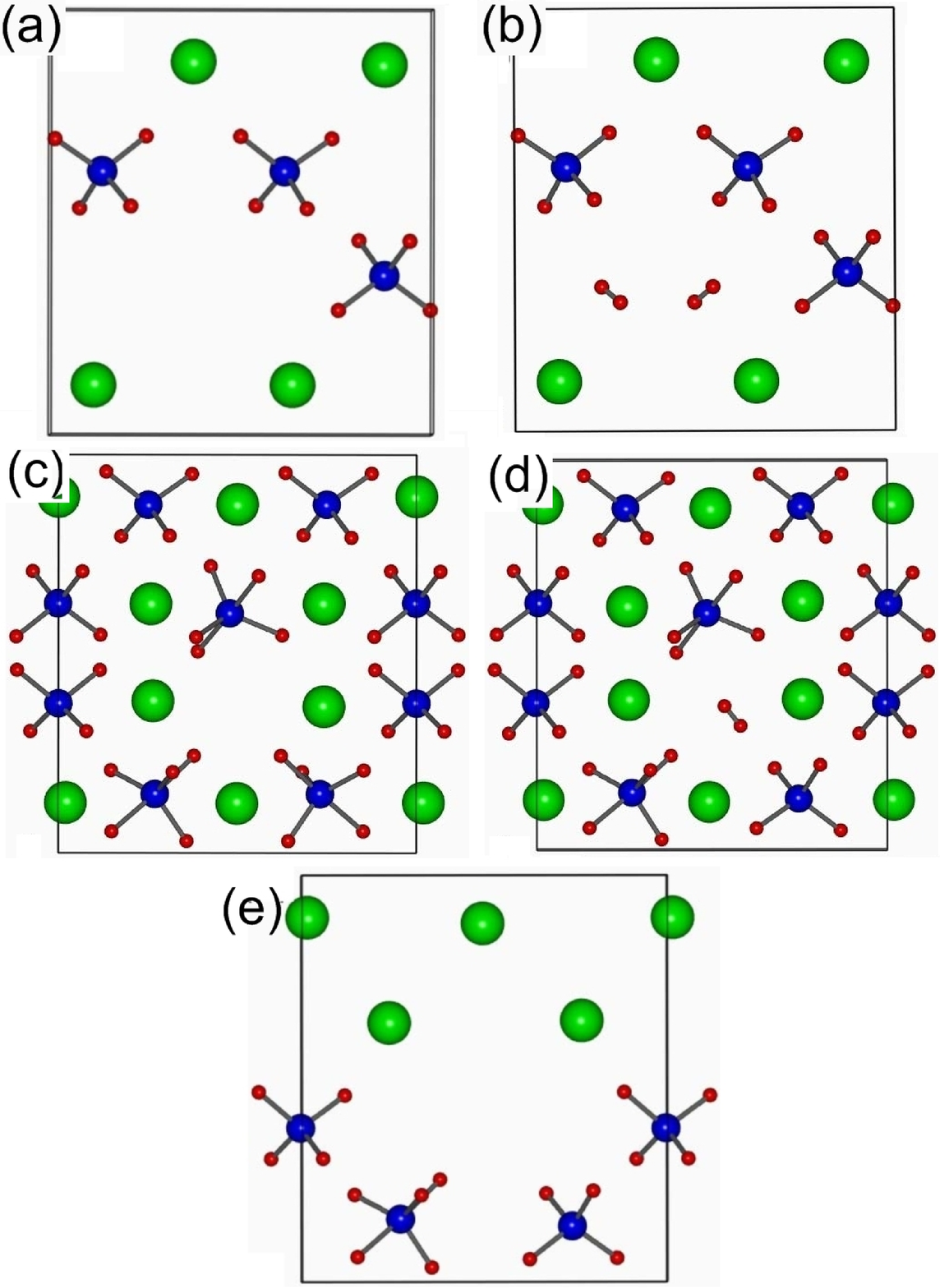}
\caption{(Color online) The structures of (a) $V_{\rm{AlH_4}}^+$, (b) $V_{\rm{Al}}^+$, (c) $V_{\rm{Al}}^{3-}$, (d) $V_{\rm{Al}}^-$, and (e) $V_{\rm{AlH_3}}$. The structural diagrams in panels (a), (b), and (e) are of sodium and aluminum atoms in a [010] plane (and coordinated hydrogens). For the cases of and $V_{\rm{Al}}^{3-}$ (c) and $V_{\rm{Al}}^-$ (d) , we show the atoms from two [010] planes in order to illustrate the differences upon reduction of the aluminum vacancy.\label{def-al}}
\end{figure}

\begin{figure}
\includegraphics[width=6in]{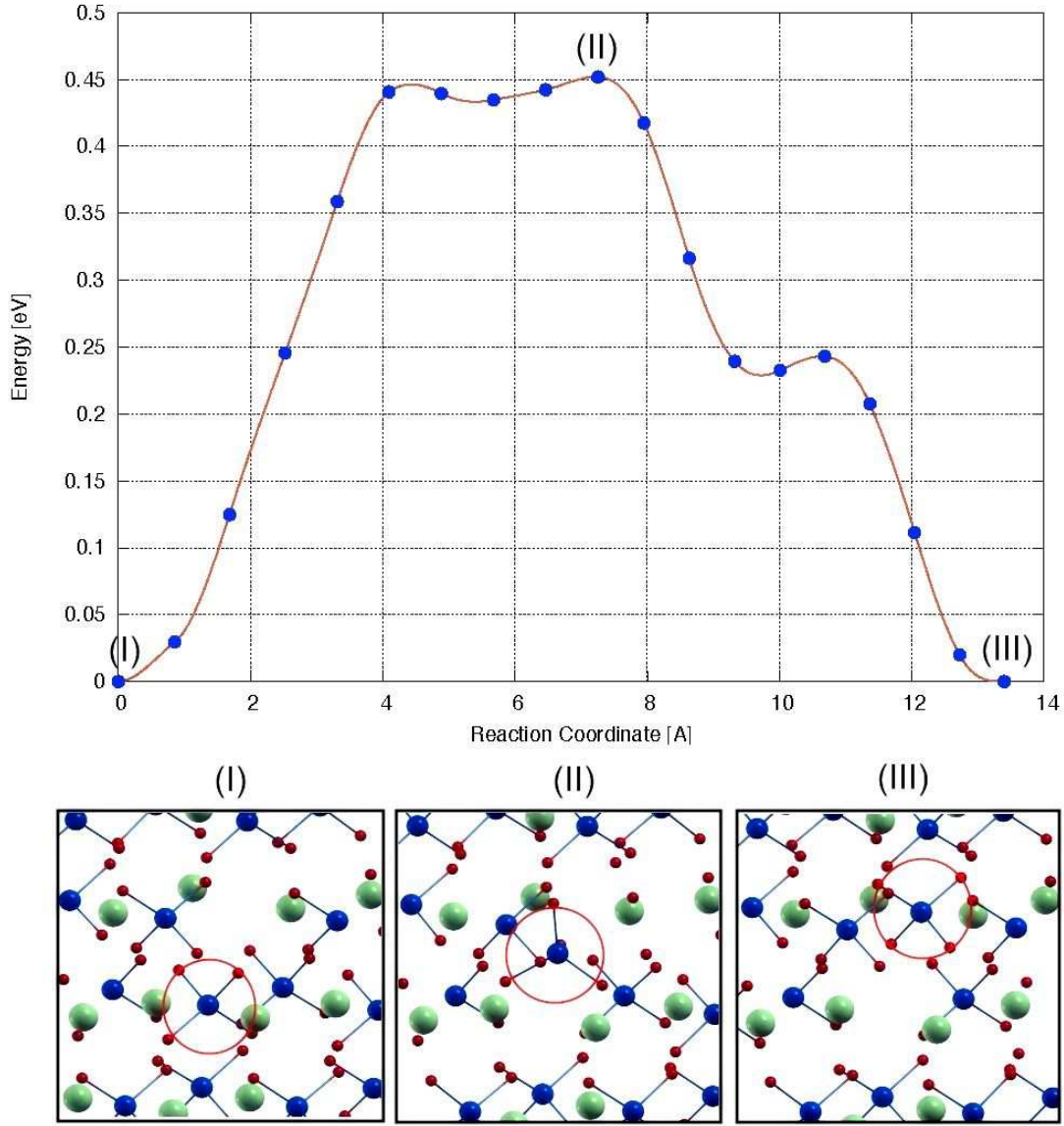}
%\vspace{-1.0in}
\caption{Energy versus reaction coordinate along the migration path for $V_{\rm{AlH_4}}^+$ (equivalently, AlH$_{4}^{-}$) as modeled with the NEB method. The local lattice structures of the initial, saddle-point, and final configurations are shown as (I), (II) and (III), respectively.
\label{fig:neb}}
\end{figure}

$V_{\rm{AlH_4}}^+$ corresponds to the removal of an entire (AlH$_4$)$^-$ tetrahedron from the system.
Figure~\ref{def-al}(a) shows that there is remarkably little structural relaxation around this defect. Figure~\ref{Aldef} shows that the formation energy of this defect is quite low, and we will see that it plays an important role. NEB calculations for the migration path are presented in Fig.~\ref{fig:neb}. Migration consists of moving a neighboring tetrahedron into the vacancy. This tetrahedron remains fairly rigid along the migration path; see Fig.~\ref{fig:neb} for snapshots of the initial, saddle-point, and final configurations and Ref.~21 for an animation of the diffusion. This path results in a migration barrier of 0.46 eV.

$V_{\rm{Al}}^+$ [Fig.~\ref{def-al}(b)] can be understood as a complex of a $V_{\rm{AlH_4}}^+$ and two (H$_2$)$_i$. The energy of $V_{\rm{Al}}^+$ is lower than the sum of its components by 1.52 eV. As discussed earlier in the case of H$_i^+$, this can be explained on the basis of the unfavorable sterics of (H$_2$)$_i$ in the absence of a vacancy.
Indeed, we note that the 1.52 eV binding energy of $V_{\rm{Al}}^+$, which contains two (H$_2$)$_i$, is very close to twice the binding energy (0.71 eV) calculated for H$_i^+$, which contains one (H$_2$)$_i$ species.

$V_{\rm{Al}}^{3-}$ is a complex of $V_{\rm{AlH_4}}^+$ and four H$_i^-$, as can be seen in Fig.~\ref{def-al}(c). The binding energy of $V_{\rm{Al}}^{3-}$, relative to the 5 component defects, is 1.46 eV. Even though the H$_i^-$ component defects in $V_{\rm{Al}}^{3-}$ all have the same charge, Coulomb repulsion between them does not seem to play a major role. The fairly large binding energy of the $V_{\rm{Al}}^{3-}$ complex can again be attributed to sterics.

$V_{\rm{Al}}^-$ can be understood as a complex of $V_{\rm{AlH_4}}^+$, (H$_2$)$_i$, and two H$_i^-$. The binding energy is 1.75 eV. The structure of $V_{\rm{Al}}^-$ is presented in Fig.~\ref{def-al}(d).

$V_{\rm{AlH_3}}$ [Fig.~\ref{def-al}(e)], finally, can be regarded as a complex of $V_{\rm{AlH_4}}^+$ and H$_i^-$. The binding energy, i.e. the difference in energy with respect to the sum of formation energies of these two constituents, is 0.37 eV. This binding energy is roughly one-quarter the binding energy of $V_{\rm{Al}}^{3-}$, as one might expect -- only one H$_i^-$ is involved in $V_{\rm{AlH_3}}$, whereas four H$_i^-$ are used to construct $V_{\rm{Al}}^{3-}$. This again implies that Coulomb repulsion between H$_i^-$ does not play an important role, and that the binding energies that are gained when defects are brought together are primarily due to sterics.

For completeness, we have also performed calculations for Al interstitials. We find, however, that these defects all have formation energies that are significantly higher than the vacancy defects discussed above. The most favorable charge state is Al$_i^{3+}$, which has a formation energy of 4.18 eV at $\epsilon_F$=3.26 eV. This value is very high and therefore the defect is not included in Fig.~\ref{Aldef}.

\subsection{Sodium-related defects}

\begin{figure}
\includegraphics[width=6in]{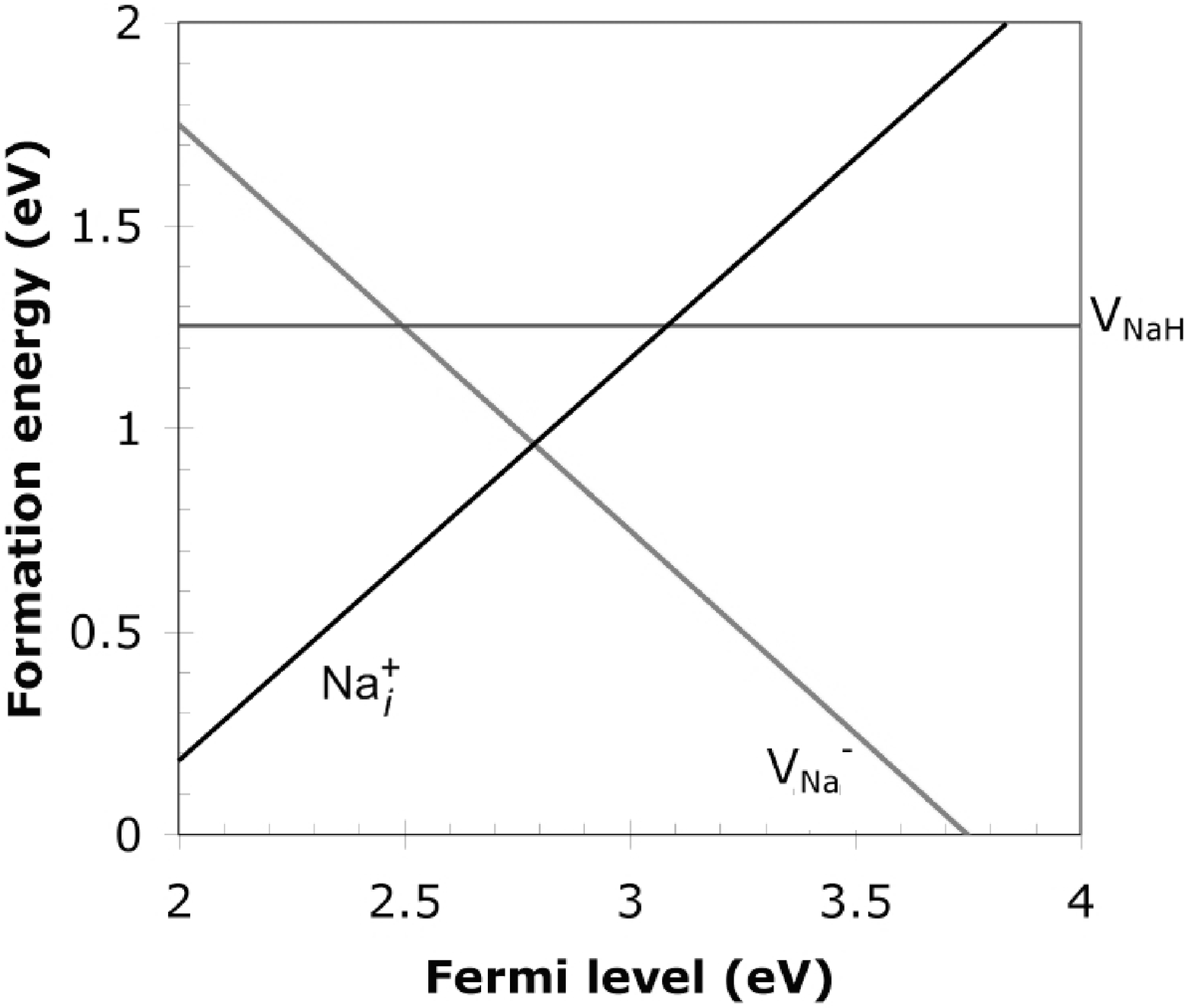}
\caption{Formation energy of sodium-related native defects in NaAlH$_4$, plotted as a function of Fermi level with respect to the valence-band maximum. Atomic chemical potentials were chosen to reflect equilibrium with Na$_3$AlH$_6$, NaAlH$_4$, and Al. \label{Nadef}}
\end{figure}

\begin{figure}
\includegraphics[width=6in]{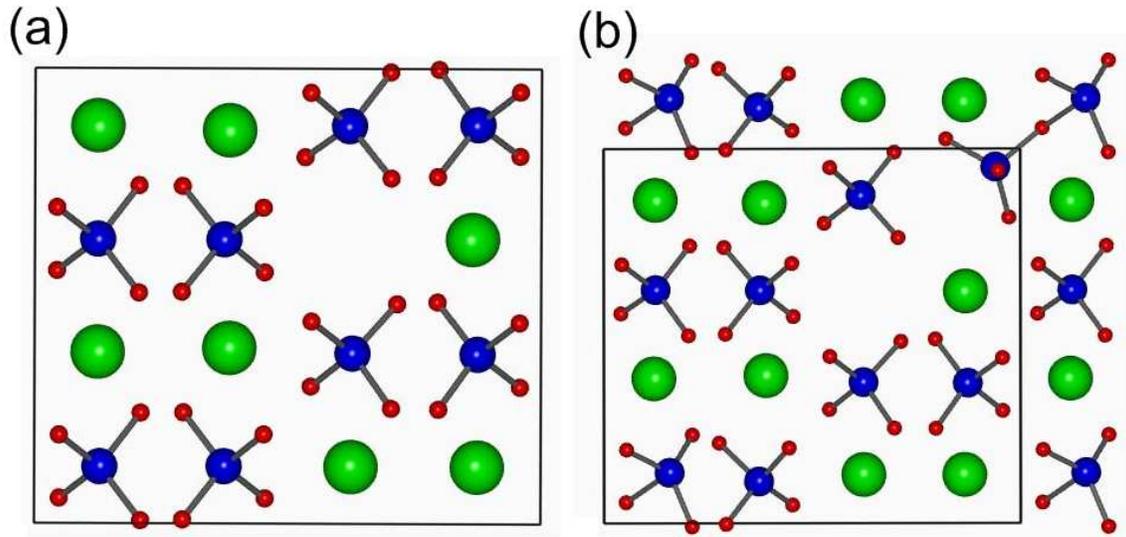}
\caption{(Color online) The structures of (a) $V_{\rm{Na}}^-$ and (b) $V_{\rm{NaH}}$. The structural diagrams are of sodium and aluminum atoms in several [010] planes (and coordinated hydrogens). \label{def-na}}
\end{figure}

Formation energies for Na-related native defects are presented in Fig.~\ref{Nadef}. $V_{\rm{Na}}^-$ corresponds to the removal of a Na$^+$ ion from the system. As in the case of $V_{\rm{AlH_4}}^+$, there is remarkably little structural
relaxation, as can be seen in Fig.~\ref{def-na}(a). We calculated a migration energy of 0.41 eV for this defect.

$V_{\rm{NaH}}$ is a complex of the defects $V_{\rm{Na}}^-$ and $V_{\rm{H}}^+$, as can be seen in Fig.~\ref{def-na}(b). The calculated binding energy is 0.14 eV. This relatively small binding energy is to be expected, since both component defects are vacancies and little energy can be gained due to strain relaxation.

We have also investigated Na interstitials. Not surprisingly, Na$_i^+$ is most favorable; it has a formation energy of 1.44 eV at $\epsilon_F$=3.26 eV and a migration barrier of 0.48 eV. Still, Na interstitials are less favorable than other defects.

\section{Discussion}
\label{sec:Discussion}

\subsection{Comparison with previous calculations}

In Sec.~\ref{sec:results}, we presented and compared formation energies of various native defects. In addition to the results presented in the figures, we also summarize key information for relevant defects in Table~\ref{tab:sum}.
We noted that interstitial defects create local stress, and that this stress can be relieved by joining interstitials and vacancies. The binding energies between such defects could be explained by this sterics argument.

{\L}odziana {\it et al.}~\cite{lodziana} have reported first-principles calculations for native point defects in NaAlH$_4$ and LiBH$_4$, using a methodology very similar to ours, though with a different choice of GGA exchange and correlation potential. Our own tests have indicated that the choice of GGA potential has only a minor impact on the results. {\L}odziana {\it et al.}~also make a different choice of chemical potentials for presenting formation energies, invoking equilibrium with Al$_2$H$_6$, but again this should affect the calculated formation energies only by a few 0.1 eV compared to our choice. Another difference is the use of supercell-size corrections for charged defects. We feel it is better not to apply uncontrolled approximations in an attempt to correct for such effects, since such corrections often tend to ``overshoot'' and make things worse rather than better.~\cite{vdw-rev-04} Although there is broad qualitative agreement between our results and those of Ref.~ \onlinecite{lodziana}, a number of notable differences appear. For instance, the transition level between the + and $-$ charge states of hydrogen vacancies and hydrogen interstitials occurs at much lower values in the band gap than it does in our calculations; the difference seems comparable to the energy of the valence-band maximum (highest occupied eigenstate) in the bulk (~$-$0.9 eV), which could indicate an error in the reference energy for the Fermi level. {\L}odziana {\it et al.}~also find a significantly higher energy for $V_{\rm{AlH_3}}^0$ and $V_{\rm{AlH_4}}^+$, even after taking the different assumptions noted above into account. Our efforts to reconcile the differences have been unsuccessful.

Gunaydin {\it et al.}~have performed first-principles calculations for neutral $V_{\rm{AlH_3}}$ and $V_{\rm{NaH}}$ in a 192-atom NaAlH$_4$ supercell.~\cite{08-ozolins-pnas} Their calculated formation energies are in reasonably good agreement with our present values, for comparable choices of chemical potentials. They obtained free-energy barriers for diffusion using an umbrella-sampling technique, resulting in values of 0.12 eV for $V_{\rm{AlH_3}}$ and 0.26 eV for $V_{\rm{NaH}}$. We feel that these computed free-energy differences are not representative of actual diffusion activation energies, which should only include enthalpy differences and do not depend on entropy terms in the free energy. This explains the difference between the values reported by Gunaydin {\it et al.}~and our values listed in Table~\ref{tab:sum}.

We also note that Gunaydin {\it et al.}~\cite{08-ozolins-pnas} seem to have scaled their calculated formation
energies and migration barriers by a factor 2/3, presumably to convert from units of ``kJ/mol'' to ``kJ/mol H$_2$'' because in Eq.~(3) of their paper 3/2 H$_2$ is being produced. Such scaling is unwarranted and unjustified, in our opinion. The formation energies and migration barriers enter into activation energies, which do not scale with the size of the system. Put differently: producing a larger amount of H$_2$ requires a given process to happen a larger number of times, but it does not increase the activation energy of that process. In the absence of this scaling factor, i.e., looking at their directly calculated numbers, the conclusions of Gunaydin {\it et al.}~are not supported by their calculations.

\subsection{Migration barriers}
\label{sec:migration}

As reported in Sec~\ref{sec:results}, migration energies were explicitly calculated in the cases of $V_{\rm{H}}^-$, $V_{\rm{H}}^+$, H$_i^-$, $V_{\rm{Na}}^-$, Na$_{i}^+$, $V_{\rm{AlH_4}}^+$, and (H$_2$)$_i$. Explicit calculations for migration paths proved very cumbersome in other cases, specifically H$_i^+$, $V_{\rm{NaH}}$, $V_{\rm{AlH_3}}$, $V_{\rm{Al}}^+$, and $V_{\rm{Al}}^-$. The reason for the complications is that these defects can be thought of as complexes between more elementary native point defects, as discussed in Sec.~\ref{sec:results}. In this view, however, one can also estimate a lower bound for the migration energies of these complexes by taking the higher of the migration barriers of the constituent defects. For instance, H$_i^+$ can be considered as a complex of $V_{\rm{H}}^+$ and (H$_2$)$_i$, suggesting that its migration barrier will be at least 0.26 eV, the value for $V_{\rm{H}}^+$. This estimate should be a lower limit for the migration barrier of the complex, since it assumes that during the migration process the complex remains fully bound. If during migration (partial) breaking of the complex occurs, then an energy cost needs to be paid that would increase the value of the barrier. Values for other defects that can be regarded as complexes were obtained in similar fashion: the aluminum-related vacancies can be regarded as combinations of $V_{\rm{AlH_4}}^+$, (H$_2$)$_i$, and H$_i^-$. $V_{\rm{NaH}}$ is regarded as a combination of $V_{\rm{Na}}^-$ and $V_{\rm H}^+$.

\subsection{Reaction mechanisms and activation energies}

A range of apparent activation energies for desorption in transition-metal-doped NaAlH$_4$ has been reported, with values as low as 0.8 eV.~\cite{02-sandrock-exp, 03-kiyobayashi-exp, 04-gross-exp, 05-kircher-exp, 07-fichtner-exp} This value is $\sim$0.4 eV lower than in undoped material.~\cite{02-sandrock-exp} The desorption process has been suggested to arise from the diffusion of defects.~\cite{00-gross-exp,07-fichtner-exp}

In order for hydrogen to desorb from this material, it is also necessary for aluminum and/or sodium to diffuse through the solid. The reaction described by Eq.~(\ref{h4-h6}) can be interpreted as the material NaAlH$_4$=(NaH)(AlH$_3$) decomposing into Na$_3$AlH$_6$=(NaH)$_3$(AlH$_3$) and AlH$_3$=(NaH)$_0$(AlH$_3$), with the latter divided into two separate phases -- Al and H$_2$. This leads us to be interested primarily in the diffusion of species corresponding to fluxes of either (NaH) or (AlH$_3$) through the material.

In Sec.~\ref{sec:chempot} we explained that for purposes of presenting our formation energy results, we chose the chemical potentials of Al, Na, and H by assuming equilibrium with NaAlH$_4$, Na$_3$AlH$_6$, and Al. Based on the calculated enthalpies of formation, and keeping in mind that the chemical potentials are referenced to the standard state of the elements, this leads to values $\mu_{\rm Na}$=$-$0.38 eV, $\mu_{\rm Al}$=0, and $\mu_{\rm H}$=$-$0.12 eV.
The resulting formation energies were presented in Figs.~\ref{Hdef}, \ref{Aldef}, and \ref{Nadef}, and in column (1) of Table~\ref{tab:sum}. Of course, other scenarios are possible. For instance, we can assume equilibrium with NaAlH$_4$, Na$_3$AlH$_6$, and H$_2$, and find $\mu_{\rm Na}$=$-$0.48 eV, $\mu_{\rm Al}$=$-$0.34 eV, and $\mu_{\rm H}$=0. The resulting formation energies are listed in column (2) of Table~\ref{tab:sum}, with the additional assumption that again the Fermi level is fixed at the value where the formation energies of $V_{\rm{AlH_4}}^+$ and $V_{\rm{Na}}^-$ are equal. As a third scenario, we can assume equilibrium with NaAlH$_4$, Al, and H$_2$, finding $\mu_{\rm Na}$=$-$0.82 eV, $\mu_{\rm Al}$=0, and $\mu_{\rm H}$=0. The resulting values are listed in column (3) of Table~\ref{tab:sum}.

We are now in a position to examine various possible mechanisms for defect-assisted diffusion and decomposition.
The reaction described by Eq.~(\ref{h4-h6}) implies formation of Na$_3$AlH$_6$ and Al during the decomposition of NaAlH$_4$. Since the formation energies depend on atomic and electronic chemical potentials, specific choices
must be made that approximate the experimental conditions as closely as possible. Our assumption is that these conditions do not differ significantly from equilibrium. In the case of charged defects, local and global charge neutrality needs to be maintained.

\subsubsection{Neutral defects}
\label{sec:neutr}

One possible mechanism involves diffusion of neutral defects such as $V_{\rm{AlH_3}}$ or $V_{\rm{NaH}}$.
The formation energy of these defects is independent of Fermi level. The total activation energy would be the sum of formation and migration energies. $V_{\rm{AlH_3}}$ and $V_{\rm{Na}}$ cannot form in the interior of the material, since this would require simultaneous formation of Al$_i$ or Na$_i$, and the total process would be much too costly.
Al- and Na-related defects will therefore necessarily be formed at an interface, where they will be directly involved in the formation of Na$_3$AlH$_6$ and/or Al. $V_{\rm{AlH_3}}$ most readily forms at an interface where NaAlH$_4$, Al and H$_2$ are in equilibrium (as also discussed in Ref.~ \onlinecite{08-ozolins-pnas}). Table~\ref{tab:sum} shows that under those conditions the activation energy for $V_{\rm{AlH_3}}$ diffusion is given by 1.13+0.46=1.59 eV. Creation of $V_{\rm{NaH}}$ would happen at an interface between NaAlH$_4$ and Na$_3$AlH$_6$, resulting in an activation energy
of 1.25+0.41=1.66 eV. These values are actually lower bounds, since as explained in Sec.~\ref{sec:migration} the migration barriers used here are only estimates. However, even these lower bounds are already higher than the observed activation energies for undoped alanate.~\cite{02-sandrock-exp} We therefore conclude it is unlikely that neutral $V_{\rm{AlH_3}}$ or $V_{\rm{NaH}}$ entities would play a dominant role.

As pointed out above, these defects are actually not elementary defects in their own right. $V_{\rm{AlH_3}}$ should be regarded as a complex consisting of $V_{\rm{AlH_4}}^+$ and H$_i^-$, while $V_{\rm{NaH}}$ is a combination of $V_{\rm{Na}}^-$ and $V_{\rm H}^+$. Indeed, based on the structure of NaAlH$_4$ it is clear that $V_{\rm{AlH_4}}^+$ and
$V_{\rm{Na}}^-$ should be regarded as the elementary Al- and Na-transporting entities in this material; note, e.g., the persistence of Al-H vibrational modes in NaAlH$_4$ up to the melting temperature.~\cite{yukawa07}

\subsubsection{Other Al- and Na-related defects}

Table~\ref{tab:sum} shows that it is too costly for defects containing only Al to diffuse. The estimated activation energies, both for $V_{\rm{Al}}$ and for Al$_i$ are simply too high. For Na-related defects, $V_{\rm{Na}}^-$ is clearly important and will be discussed in more detail below. Na$_i^+$, on the other hand, is too costly to form.

\subsubsection{Role of hydrogen-related defects}

In Sec.~\ref{sec:neutr} we noted that $V_{\rm{AlH_4}}^+$ and $V_{\rm{Na}}^-$ should be regarded as the elementary defects responsible for Al and Na transport. Note that we do not necessarily need {\it both} to explain the decomposition of NaAlH$_4$; either defect by itself would suffice to explain formation of Na$_3$AlH$_6$ and Al.
The accompanying defects (H$_i^-$ and $V_{\rm H}^+$) discussed in Sec.~\ref{sec:neutr} are hydrogen-related defects that provide local charge neutrality and additional hydrogen transport; however, there is no pressing reason why, out of possible hydrogen-related defects, H$_i^-$ should be the only one associated with $V_{\rm{AlH_4}}^+$, or $V_{\rm H}^+$ the only one associated with $V_{\rm{Na}}^-$. The formation energy of the hydrogen-related defects is sensitive to the choice of chemical potentials and the position of the Fermi level, and different conditions may favor different defects. Table~\ref{tab:sum} shows that among the positively charged defects, $V_{\rm H}^+$ is systematically lower in energy than H$_i^+$, and among the negatively charged defects, H$_i^-$ is lower in energy than $V_{\rm H}^-$. However, different Fermi-level conditions will favor either positively or negatively charged defects.

It is then important to realize that, from the point of view of transporting hydrogen, H$_i^-$ is equivalent to $V_{\rm H}^+$. Indeed, moving a hydrogen interstitial from left to right is equivalent to moving a vacancy from right to left.
And given the opposite charge on these defects, local electric fields would indeed tend to push these defects in opposite directions.

We look upon H$_i^-$ and $V_{\rm H}^+$ as charge-carrying defects that will provide the necessary local charge neutrality. In the absence of electronic charge carriers (electrons or holes), it is these highly mobile hydrogen-related defects that will provide local charge balance in an insulator such as NaAlH$_4$.

Unlike the Na- or Al-related defects, the hydrogen-related defects can readily form in the interior of the material. H$_i^-$ and $V_{\rm H}^+$ constitute a Frenkel pair that can be formed simply by moving a substitutional H atom to an interstitial position. This contrasts sharply with the Al- or Na-related defects, which cannot form in the bulk but only at an interface, as discussed in Sec.~\ref{sec:neutr}. The hydrogen-related defects are thus unique in their ability to form within the bulk, and given their modest formation energies, we expect a finite concentration of such pairs to always be present.

We suggest here that it is the formation of these hydrogen defect Frenkel pairs that may be the rate limiting step for the desorption and decomposition process. Our calculated formation energy for the H$_i$-$V_{\rm H}$ pair is 1.10 eV, corresponding to the configuration where the pair creates an AlH$_{4}$-H-AlH$_{3}$-H-AlH$_{3}$ (or Al$_{3}$H$_{12}$) complex. Note that this is lower than the sum of the formation energies of H$_i^-$ and $V_{\rm H}^+$ listed in Table~\ref{tab:sum}, the reason being that the proximity of the defects leads to a finite binding energy, which we calculate to be equal to 0.46 eV. This calculated formation energy would yield an estimate for the activation energy of formation of this pair of 1.10+0.16=1.26 eV, where we have added the migration barrier of H$_i^-$ to the formation energy as an estimate for the barrier that needs to be overcome. This activation energy is consistent with the experimentally observed desorption activation energies in undoped NaAlH$_4$.

This suggested mechanism also immediately provides an explanation for the observed effect of Ti (or other transition metals), following the arguments in Ref.~ \onlinecite{07-peles-theory2}. Ti acts as an electrically active dopant that shifts the Fermi level. Shifts of the Fermi level will always lead to a lowering of the formation energy of the hydrogen-related defects.~\cite{07-peles-theory2} For purposes of illustration, consider an upward shift in the Fermi level, such as would be introduced by Ti.~\cite{07-peles-theory2} Figure~\ref{Hdef} shows that if the magnitude of this shift exceeds a few 0.1 eV, it becomes energetically more favorable for the vacancy to be present in the negative charge state rather than the positive charge state. The H$_i^-$-$V_{\rm H}^-$ Frenkel pair will then be lower in energy than the H$_i^-$-$V_{\rm H}^+$ pair. To illustrate the effect of a shift in Fermi level, consider the example given in Ref.~\onlinecite{07-peles-theory}: an upward shift of 0.44 eV in the Fermi level leads to a decrease in the formation energy of $V_{\rm H}^-$ by 0.44 eV and an increase in its concentration [Eq.~(\ref{eq:conc})] by 6 orders of magnitude at 100$^\circ$C. We suggest that this energy gain, in the presence of an electrically active impurity that shifts the Fermi level, explains the lowering in activation energy for decomposition and desorption. Note that the components of the H$_i^-$-$V_{\rm H}^-$ pair repel each other, which eliminates any further energy cost in separating them into isolated defects.

\section{Conclusions}

We have studied native defects in NaAlH$_4$ based on first-principles density functional calculations. Despite the wide variety of possible native defects in this material, simplifying principles can be applied to qualitatively understand the relevant defects. Many of the defects can be described as complexes containing the following elementary defects in the system: $V_{\rm{AlH_4}}^+$, $V_{\rm{Na}}^-$, $V_{\rm{H}}^+$, H$_i^-$, and (H$_2$)$_i$. This does not eliminate the need for performing explicit calculations, since the interactions between constituent defects significantly lower the energy of the complex, e.g., through relaxation of local strains. We suggest that this view of native defects will also be useful when studying other alkali or alkaline earth aluminum hydrides.

Our results should be of use in interpreting experimental information on desorption from NaAlH$_4$ and the effects of dopants such as transition metal impurities on the kinetics. Charged (rather than neutral) defects play the dominant role in diffusion and decomposition. We have proposed specific mechanisms for this process; in particular, we suggest that hydrogen-related Frenkel pairs provide the charge carriers that aid in diffusion of Al- or Na-related defects. The dependence of their formation energy and, therefore, their concentration on Fermi level may explain the observed effects of transition-metal dopants. Moreover, one could engineer the hydrogen desorption kinetics by adding electrically active impurities to shift the Fermi level for tuning the concentrations of the defects relevant for hydrogen desorption. The impurities suitable for that purpose do not necessarily need to be transition metals.

\begin{acknowledgments}

This work was supported by the U.~S.~Department of Energy (Award No.: DE-FG02-07ER46434). It made use of the CNSI Computing Facility under NSF Grant No.~CHE-0321368, the Ranger supercomputer from the TeraGrid computing resources supported by the NSF under grant No. DMR070072N, and NERSC resources supported by the DOE Office of Science under Contract No. DE-AC02-05CH11231.

\end{acknowledgments}

%\clearpage
%\pagebreak

\clearpage
\pagebreak

\begin{table}
\caption{Calculated formation energies $E^f$ and migration energies $E_m$ for selected defects in NaAlH$_4$. For charged defects, the formation energies are taken at the Fermi-level position where the formation energies of $V_{\rm{AlH_4}}^+$ and $V_{\rm{Na}}^-$ are equal. (1), (2), and (3) refer to different choices of chemical potentials.
(1) corresponds to equilibrium with NaAlH$_4$, Na$_3$AlH$_6$, and Al ($\mu_{\rm Na}$=$-$0.38 eV, $\mu_{\rm Al}$=0, $\mu_{\rm H}$=$-$0.12 eV; $\epsilon_F$=3.26 eV); (2) to equilibrium with NaAlH$_4$, Na$_3$AlH$_6$, and H$_2$ ($\mu_{\rm Na}$=$-$0.48 eV, $\mu_{\rm Al}$=$-$0.34 eV, $\mu_{\rm H}$=0$; \epsilon_F$=3.15 eV); and (3) to equilibrium with NaAlH$_4$, Al, and H$_2$ ($\mu_{\rm Na}$=$-$0.82 eV, $\mu_{\rm Al}$=0, and $\mu_{\rm H}$=0; $\epsilon_F$=2.81 eV).
Conditions (1) correspond to the case depicted in Figs.~\protect\ref{Hdef}, \protect\ref{Aldef}, and \protect\ref{Nadef}. Migration energies denoted by an asterisk (*) are lower bounds estimated by considering the defect as a complex and taking the higher of the migration energies of the constituents; see text.
\label{tab:sum}}
\begin{center}
\begin{ruledtabular}
\begin{tabular}{ccccc}
Defect 				& 			& $E^f$ (eV) &			& $E_m$ (eV) \\
					&	(1)		&	(2)		&	(3)		&			\\
\tableline
$V_{\rm{H}}^+$		&	0.90	&	0.91	&	0.57	&	0.26	\\
$V_{\rm{H}}^-$		&	1.00	&	1.23	&	1.57	&	0.92	\\
H$_i^+$				&	1.27	&	1.04	&	0.70	&	0.26*	\\
H$_i^-$				&	0.66	&	0.65	&	0.99	&	0.16	\\
(H$_2$)$_i$ 		&	1.07	&	0.84	&	0.84	&	0.25	\\
$V_{\rm{Al}}^+$		& 	1.12	& 	0.67	& 	0.67	&	0.46*	\\
$V_{\rm{Al}}^-$ 	&	1.13	&	0.90	&	1.58	&	0.46*	\\
$V_{\rm{Al}}^{3-}$ 	&	1.67	&	1.66	&	3.02	&	0.46*	\\
$V_{\rm{AlH_4}}^+$	&	0.49	&	0.50	&	0.50	&	0.46	\\
$V_{\rm{AlH_3}}^0$	&	0.78	&	0.79	&	1.13	&	0.46*	\\
$V_{\rm{AlH_2}}^0$	&	1.07	&	0.96	&	1.30	&	0.46*	\\
Al$_i^{3+}$			&	4.18	&	4.19	&	2.83	&	-		\\
$V_{\rm{Na}}^-$		&	0.49	&	0.50	&	0.50	&	0.41	\\
$V_{\rm{NaH}}$		&	1.25	&	1.27	&	0.93	&	0.41*	\\
Na$_i^+$			&	1.44	&	1.43	&	1.43	&	0.48	\\
\end{tabular}
\end{ruledtabular}
\end{center}
\end{table}

\clearpage
\pagebreak

%\clearpage
%\pagebreak
\newpage

%\clearpage
%\pagebreak
\newpage

%\clearpage
%\pagebreak
\newpage

%\clearpage
%\pagebreak
\newpage

%\clearpage
%\pagebreak
\newpage

%\clearpage
%\pagebreak
\newpage

\newpage

\end{document}